# High-refractive index and mechanically cleavable non-van der Waals InGaS$_3$

A. N. Toksumakov [1, 2], G. A. Ermolaev[1], A. S. Slavich[1], N. V. Doroshina[1], E. V. Sukhanova[2], D. I. Yakubovsky [1], S. M. Novikov[1], A. S. Oreshonkov[3], D. M. Tsymbarenko[4], Z. I. Popov[2], D. G. Kvashnin[2], A. A Vyshnevyy[1], A. V. Arsenin[1, 5], D. A. Ghazaryan[1, *], and V. S. Volkov[1, 5, *]

[1]Center for Photonics and 2D Materials, Moscow Institute of Physics and Technology, Dolgoprudny 141701, Russia

[2]Emanuel Institute of Biochemical Physics RAS, Moscow 119334, Russia

[3]School of Engineering and Construction, Siberian Federal University, Krasnoyarsk 660041, Russia

[4]Department of Chemistry, Lomonosov Moscow State University, Moscow 119991, Russia

[5]XPANCEO, Moscow 127495, Russia

[*]Authors to whom correspondence should be addressed: kazarian.da@mipt.ru and volkov.vs@mipt.ru

**Abstract**

The growing families of two-dimensional crystals derived from naturally occurring van der Waals materials offer an unprecedented platform to investigate elusive physical phenomena and could be of use in a diverse range of devices. Of particular interest are recently reported atomic sheets of non-van der Waals materials, which could allow a better comprehension of the nature of structural bonds and increase the functionality of prospective heterostructures. Here, we study the optostructural properties of ultrathin non-van der Waals InGaS$_3$ sheets produced by standard mechanical cleavage. Our ab initio calculation results suggest an emergence of authentically delicate out-of-plane covalent bonds within its unit cell, and, as a consequence, an artificial generation of layered structure within the material. Those yield to singular layer isolation energies of ~ 50 meVÅ$^{-2}$, which is comparable with the conventional van der Waals material's monolayer isolation energies of 20 - 60 meVÅ$^{-2}$. In addition, we provide a comprehensive analysis of the structural, vibrational, and optical properties of the materials presenting that it is a wide bandgap (2.73 eV) semiconductor with a high-refractive index (> 2.5) and negligible losses in the visible and infrared spectral ranges. It makes it a perfect candidate for further establishment of visible-range all-dielectric nanophotonics.

**Introduction**

Research on layered materials with relatively weak out-of-plane van der Waals bonds is of great significance nowadays. These layered materials are studied in natural[1,2] and artificial heteroformations[3–6]. Accounting for their stoichiometric composition and crystal structure, one can classify the former into graphene, transition metal dichalcogenide or mono- chalcogenide, sulfosalt, oxide, neo-, phyllo- silicate and phosphate families. The latter is assembled physically by stacking atomic layers with strong in-plane covalent bonds of the former on top of each other[3,7]. In addition, one can also introduce a nominal categorization for such layered materials based on their cleavage feasibility into separate atomic sheets[8]. It could be understood as an energy required for the isolation or separation of a singular atomic layer from the bulk[9,10]. In first-principles calculations, the interlayer binding energy per unit area is a key parameter that determines the probability of such an event. Within this formalism, it is reasonable to distinguish naturally, potentially cleavable, and robust materials with threshold exfoliation energies of the order of magnitudes of[8] 10 meVÅ$^{-2}$, 100 meVÅ$^{-2}$, and 1000 meVÅ$^{-2}$, respectively. Apart from natural van der Waals formations, some of the cleavable materials lack the out-of-plane van der Waals bonds in their structure. They form interlayer bonds of a different nature, e.g., covalent, but of comparable strengths and bear the name of non-van der Waals materials[11–18]. Vivid examples attained on probation are atomic sheets of $Fe_2O_3$[15], $AgCrS_2$[16] and $FeS_2$[19], derived through sonication-assisted and cation-intercalation exfoliation methods, respectively.

In this article, we explore the optostructural properties of the hexagonal phase of $InGaS_3$ expanding the list of non-van der Waals materials. We demonstrate that unlike orthorhombic $InGaS_3$ predicted theoretically[20], its hexagonal phase obeys a non-layered arrangement of crystal structure. Peculiarly, it can be cleaved down to atomically thin sheets due to the presence of few and delicate out-of-plane bonds within its unit cell. Our findings are based on systematic study of the crystal lattice through X-ray crystallography and supported by first-principles calculations. In addition, we demonstrate an alternative crystal structure reckoning technique combining the spectroscopic ellipsometry with first-principles calculations. It allows us to resolve that the hexagonal $InGaS_3$ is a wide bandgap (2.73 eV) semiconductor with high-refractive index (> 2.5) and negligible losses for visible and infrared spectra, which makes it relevant for next-generation nanophotonics and mietronics.

**Results**

**Structure and morphology of non-van der Waals $InGaS_3$**

Typical crystals of $InGaS_3$ appear with yellow-to-lustrous grey shades as displayed in Figure 1 (a). Such colours are determined by the 1:1:3 ratio distribution of Indium: Gallium: Sulphur atoms (see energy-dispersive X-ray spectroscopy pattern in Supplementary Figure S1). The crystals exhibit a hexagonal arrangement of III-III-IV group elements in the $P6_5$ space group with lattice parameters of $a = b = 6.6$ Å and $c = 17.9$ Å as presented in the inset of Figure 1 (a). Our all-through crystallographic imaging results confirm its hexagonal structure. Figure 1 (b and c) demonstrate the selected X-ray diffraction patterns across two eminent $a*b*$ and $c*b*$ crystallographic planes. Notably, a similar hexagonal structure of $InGaS_3$ was previously reported with a high *R*-value, 8.1 %, in the chiral $P6_1$ space group[21]. However, testing the space group for the dataset collected from our samples resulted in an unsatisfactory Flack's parameter of 0.94 (3) imposing a requirement for swapping within an enantiomorphic pair, $P6_1$ -> $P6_5$[22]. Further details on data collection, refinement parameters and interatomic distances are provided in Methods and Supplementary Tables ST1, ST2. In addition, the hexagonal structure of the material was alternatively verified through transmission electron microscopy (see Supplementary Figure S2 for diffraction pattern across $a*b*$ crystallographic plane reflex interpretations).

The hexagonal $InGaS_3$ contains various structural bonds, whose strengths were estimated by first-principles calculations based on density functional theory (DFT). To obtain energies required for the isolation of individual atomic sheets, we estimated the differences among the ground-state energy of the relaxed structure and all of its unrelaxed states. Afterwards, we looked for the planes with the minimal binding energies to determine potentially breakable, or in our case, cleavable directions. Excluding relaxation energies along the *c*-axis, we found the exfoliation energy of $E_{exf} \approx 53$ meVÅ$^{-2}$ for the planes shown in Figure 1 (d). After accounting for the further relaxation of such isolated layers[23], we obtained an even smaller exfoliation energy of $E^*_{exf} \approx 21$ meVÅ$^{-2}$. Figure 1 (e) compares our results with the evaluated exfoliation energies of conventional van der Waals (and non-van der Waals) materials. Notably, the energies of conventional van der Waals materials cover the range of 20 to 60 meVÅ$^{-2}$, unlike for non-van der Waals ones with a larger dispersion in the range of 25 to 180 meVÅ$^{-2}$. Our InGaAs$_3$ has relatively low exfoliation energy among other non-van der Waals materials, which is slightly lower than that of some van der Waals materials, such as PdS$_2$ and PdSe$_2$.

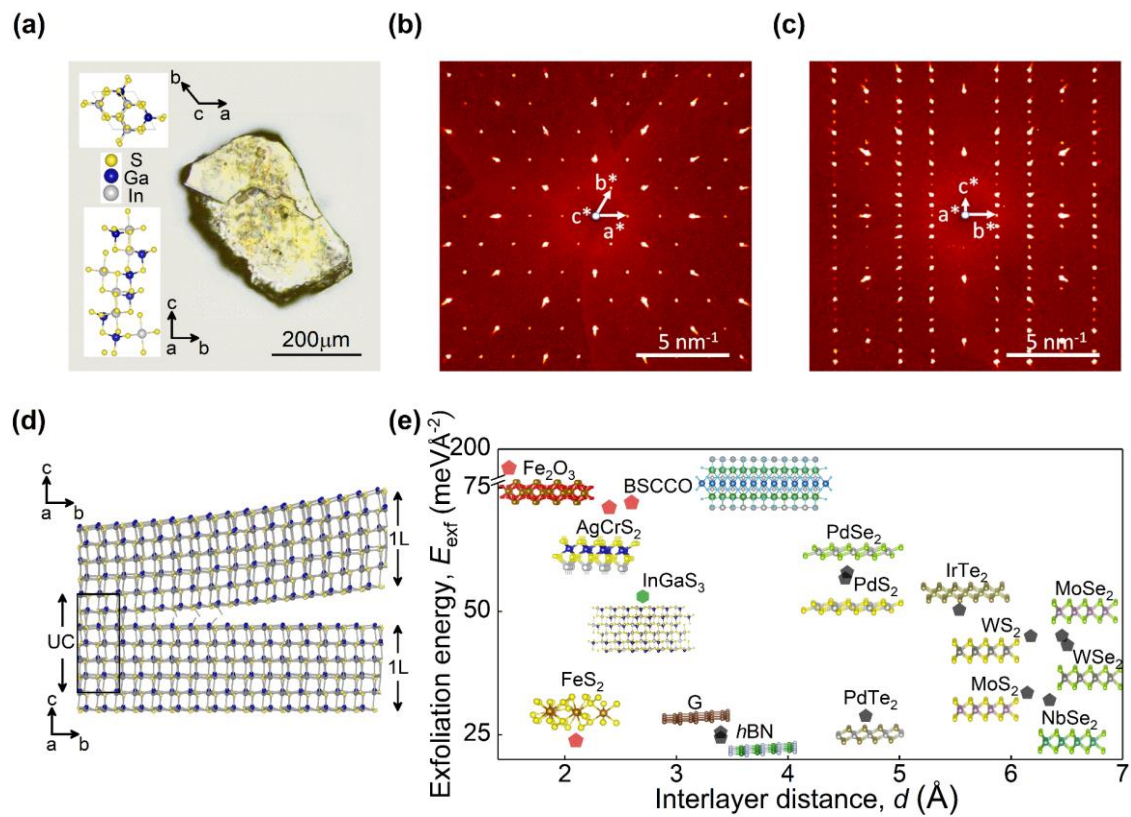

**Figure 1: Crystal structure and interplane binding energies of non-van der Waals InGaS$_3$. a** 5X optical micrograph of bulk crystal on a glass slide. Inset: Top and side views of P6$_5$ space group hexagonal crystal structure with lattice constants of $a = b = 6.6$ Å, $c = 17.9$ Å. The *a*, *b*, *c* and *a\**, *b\**, *c\** correspondingly introduce real and reciprocal space axes throughout the figure. **b** X-ray diffraction micrograph of the crystal presented in panel (a) across *a\*b\**-plane. **c** Same as (b), but across *c\*b\**-plane. **d** Schematic representation of artificially generated layered structure achieved by cutting off the delicate non-van der Waals bonds. UC stands for unit cell, 1L for artificially created monolayer. **e** Typical exfoliation energies of conventional van der Waals (black pentagons) and non-van der Waals (red pentagons) materials evaluated by first-principles calculations. Green hexagon presents our evaluation of the exfoliation energy along the proposed minimal energy atomic plane.

Slightly elevated temperature treatment within the standard exfoliation procedure[7] allows the cleavage of InGaS$_3$ crystals into individual nanosheets with smooth surfaces (see Figure 2(a)). In contrast, the ultrathin atomic sheets preserve their flatness to a certain extent (see Supplementary Figure S3 for atomic-force micrographs of inhomogeneous atomic sheets). This is attributed to the lack of apparent layered structure with out-of-plane van der Waals bonds within the material. We presume that the inhomogeneous sheets are obtained owing to the arbitrary breakage of some of the other, less delicate, non-van der Waals bonds within the material. Figure 2 (b-d) show AFM scans of pristine atomic sheets with nearly atomically flat surfaces (root-mean-square roughness of 0.3 nm). Thicknesses of these sheets correspond to one (1L), two (2L), three (3L), four (4L) and five (5L) artificially generated monolayers.

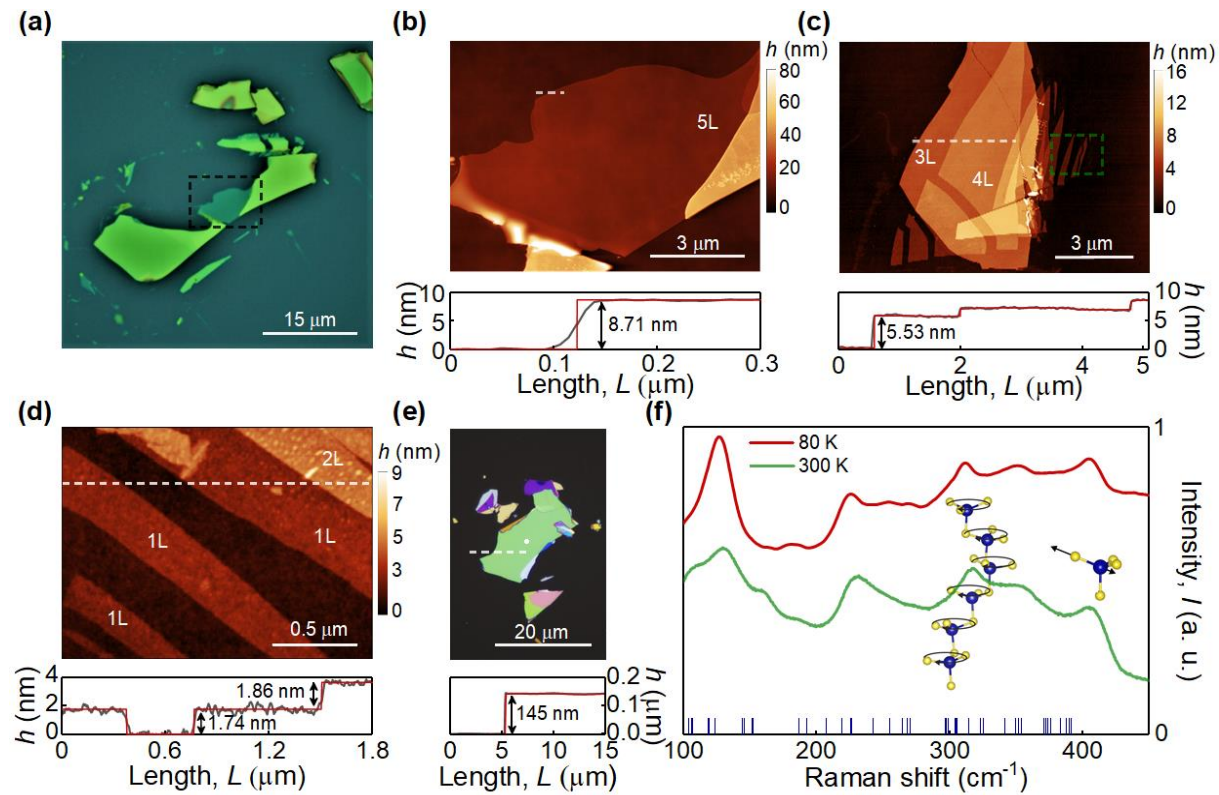

**Figure 2: Cleavage feasibility of non-van der Waals InGaS$_3$. a** 100X optical micrograph of a typical nanosheet on Si/SiO$_2$ substrate. **b** AFM topographical scan of atomic sheets over the region marked by dashed rectangle in (a). Height profiles are taken along dashed lines throughout the figure, thickness equivalent to 5L (8.71±0.38 nm). **c** Same as (b), but for thinner atomic sheets, thickness equivalent to 3L (5.53±0.32 nm) and 4L (7.3±0.34nm). **d** Same as (c), but a magnified region marked by a dashed rectangle in panel (c) and for atomic strips instead, thickness equivalent to 1L (1.74±0.24nm) and 2L (1.86±0.28nm). **e** 100X optical micrograph of typical nanosheet on a quartz substrate. White dot marks the Raman spectra collection point (145±4nm). **f** Raman spectra of the nanosheet presented in panel (e) at room (green line) and liquid nitrogen (red line) temperatures. Vertical bars label evaluated Raman-active modes (see Supplementary ST3 for details). Insets: Schematic representation of lattice vibration directions and corresponding Raman-active modes.

Figure 2 (f) shows Raman spectra of hexagonal InGaS$_3$ nanosheets performed at room and liquid nitrogen temperatures. Obtained spectral profiles have a complex form due to the overlapping bands, although peaks are slightly sharper at lower temperatures. The profiles display deeps at wavenumbers of 180 - 200 cm$^{-1}$ and 270 - 290 cm$^{-1}$, which is in a good agreement with our lattice dynamics calculations (see vertical bars on the inset of Figure 2 (f)). Those predict an absence of vibrational modes within these wavenumber ranges (see Methods for calculation details). For hexagonal InGaS$_3$ in the P6$_5$ space group, there are 90 normal vibrational modes in the center of the first Brillouin zone. Those can be represented by the following expression $\Gamma_v$ = 15*A* + 15*B* + 15*E$_1$* + 15*E$_2$*. Here, *A* + *E$_1$* are acoustical translational modes, *B* are the silent modes, 14A + 14*E$_1$* are infrared-active modes, while 14*A* + 14*E$_1$* + 15*E$_2$* are Raman-active modes. Their wavenumbers at zero-pressure are presented in Supplementary Table ST3. The contribution of individual atoms to vibrational modes - the partial phonon density of states (pPDoS), is presented in Supplementary Figure S4. According to our pPDoS evaluations, the strong peak at 125 cm$^{-1}$ is a mixed vibration of Gallium and Indium atoms, while the weak shoulder at the low-wavenumber region (157 cm$^{-1}$) is related to Gallium translations. The middle range of the spectra (from 200 to 270 cm$^{-1}$) associates to Sulphur vibrations. The high-wavenumber spectral part, above 290 cm$^{-1}$, mostly associates with Ga-S vibrations. An example of Gallium-Sulphur stretching-like mode is shown in Figure 2 (f, right inset). For the tetrahedral GaS$_4$ combined in chains through common Sulphur atoms emerge specific rotational vibrations of structural units, for instance, those that are demonstrated in Figure 2 (f, mid inset). In this case, the rotation of triangle-like GaS$_3$ units is observed. Raman spectra of thinner, up to bilayer, sheets are provided in Supplementary Figure S5. Those show no apparent shift in peak positions and a generally expected decline in their amplitudes.

**Anisotropic optical properties and crystal structure reckoning of non-van der Waals InGaS$_3$**

We investigated the anisotropic dielectric tensor (Figure 3(a)) of hexagonal InGaS$_3$ through imaging spectroscopic ellipsometry in the tiny region of interest of 10 μm$^2$ within the area of exfoliated sheets. The ability to focus on the homogeneous high-quality region is a key benefit of our imaging technology[24].

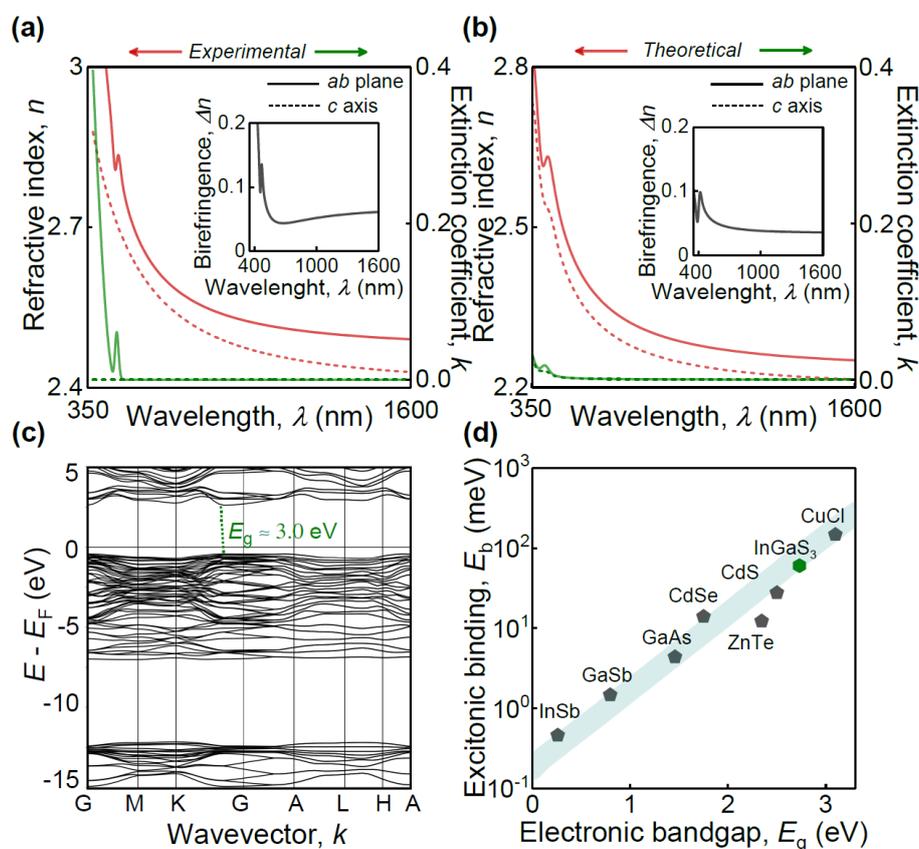

Figure 3: Non-van der Waals InGaS$_3$ optical response. **a** Experimental and **b** first-principle refractive indices and extinction coefficients for *ab*-plane (straight line) and *c*-axis (dashed line) of commensurate values and trends validating the hexagonal structure of InGaS$_3$ (see Figure 1(a (inset) and d)). Insets: material's birefringence. **c** Electronic bandstructure from first-principle calculations. Fermi energy ($E_F$) is shifted to zero for clarity. Electronic bandgap is of $E_g$ = 3 eV. **d** Excitonic binding energy versus electronic bandgap in traditional semiconductor materials

Furthermore, we simultaneously recorded and analysed optical responses from four InGaS$_3$ sheets to ensure a great precision and reproducibility of our results (see Methods and Supplementary Information for details). The dielectric function of hexagonal InGaS$_3$, similarly to transition metal dichalcogenides, is best represented by Tauc-Lorentz oscillators[25,26] and Cauchy model[27] across the crystallographic *ab*-plane and *c*-axis, respectively. Despite this, the non-van der Waals interaction reduces InGaS$_3$ birefringence to a relatively small value of Δ*n* ~ 0.1 (the inset in Figure 3(a)) in contrast to the huge anisotropy of Δ*n* ~ 1.5 in transition metal dichalcogenides with natural van der Waals bonds[24]. To explain this relatively weak anisotropy, we have performed additional DFT calculations (see Figure 3 (b)). The comparison of plots in Figure 3 (a and b) reveals a good agreement between experimental and theoretical results. Furthermore, our evaluations confirm the crystal structure of InGaS$_3$ (see inset of Figure 1(a)) since dielectric response is a fingerprint of the material's electronic bandstructure (see Supplementary Information for details). We also evaluated optical constants for the orthorhombic phase of InGaS$_3$ predicted theoretically in recent work[20]. Obtained constants are presented in Supplementary Figure S7. Those fail to successfully reproduce obtained experimental results. Hence, our technique, which combines a spectroscopic ellipsometry with density functional calculations unambiguously confirms the hexagonal structure of InGaS$_3$ in addition to the X-ray analysis, and can be potentially used for the identification of crystal structures of other materials.

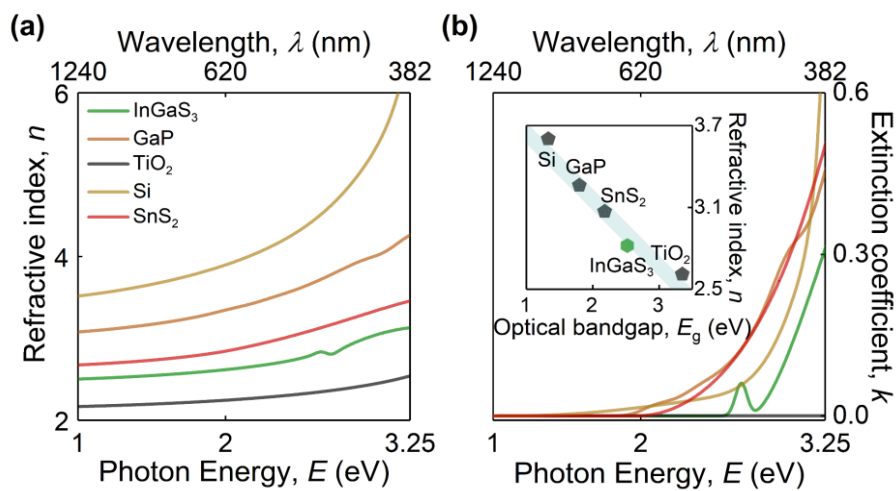

**Figure 4: Towards visible-range all-dielectric nanophotonics with non-van der Waals InGaS$_3$. a** Refractive index and **b** extinction coefficient spectra of visible-range high-refractive index and low-loss materials. Inset: The correlation between refractive indices and optical bandgaps for high-refractive index materials.

Apart from optical constants, we also evaluated the electronic bandstructure of hexagonal InGaS$_3$ (Figure 3(c)) and individual orbital-resolved density of electronic states (see Supplementary Figure S7) using the Heyd-Scuseria-Ernzerhof hybrid functional (see Methods)[28]. For the latter, the main contributions to the bottom of conduction bands come from *s*-states of Gallium, Indium, and *p*-states of Sulphur atoms. From Figure 3(c), we conclude that InGaS$_3$ is an indirect bandgap semiconductor with $E_g$ = 3.0 eV, which is close to bandgap ($E_g$ = 2.73 eV) obtained experimentally from ellipsometry analysis. Moreover, we observed an excitonic peak at $E_{exc}$ = 2.67 eV (464 nm) in optical response of the material, which is shown in Figure 3 (a). Its excitonic binding energy is $E_b = E_g - E_{exc}$ = 60 meV. Interestingly, these values ($E_g$ = 2.73 eV and $E_b$ = 60 meV) are in line with bandgaps and excitonic binding energies of traditional semiconductors (Figure 3(d)). This indicates that the exciton in InGaS$_3$ has the same physical origin and thus can be described by Wannier-Mott model[29].

Finally, we would like to highlight the strong dielectric response of InGaS$_3$ (Figure 4(a)). Its refractive index is comparable to indices of classical high-index materials, such as silicon (Si)[30] and titanium oxide (TiO$_2$)[31,32], as well as recently emerging high-index materials, such as Gallium Phosphide (GaP)[33] and tin(IV) Sulphide (SnS$_2$)[34], as shown in Figure 4(a). More importantly, InGaS$_3$ has zero optical losses up to 2.57 eV (485 nm), as can be seen in Figure 4(b). This value lies between optical bandgaps of SnS$_2$ ($E_g$ = 2.18 eV) and TiO$_2$ ($E_g$ = 3.35 eV). Hence, InGaS$_3$ covers an important gap in visible spectral range for high-refractive index materials (see inset in Figure 4(b)). As a result, we conclude that it is an outstanding material for the novel research direction of all-dielectric meta-optics in the visible range[33,35].

**Discussion**

To sum up, we demonstrated that hexagonal InGaS$_3$ can be cleaved down to thicknesses of artificially created individual monolayers due to delicate and weak non-van der Waals bonds within its unit cell. It is made possible due to a specific arrangement of atoms across the crystallographic *c*-axis, where the exfoliation energy diminishes to 53 meVÅ$^{-2}$ for a selected atomic plane. This exfoliation energy value is within the range of conventional van der Waals materials. Furthermore, the hexagonal InGaS$_3$ is well suited for next-generation nanophotonics, mie-tronics[36], and expands high-refractive index pallette[37]. In particular, it meets two major requirements of mietronics: high-refractive index ($n$ > 2.5) and broadband transparency (k ≈ 0 above 465 nm). Its refractive index is about ~10% higher than of traditional TiO$_2$[38]. Hence, all-dielectric nanostructures, such as waveguides[39], Mie-resonance nanoparticles[40], and subwavelength metasurfaces[41] based on InGaS$_3$ should be ~10% more effective and compact. Additionally, it exhibits an out-of-plane optical anisotropy (Δ$n$ ~ 0.1), in contrast to conventional high-refractive index materials, which may greatly extend its scope of applications. Therefore, InGaS$_3$ possesses a unique combination of optical properties, such as high-refractive index, zero optical losses and out-of-plane anisotropy also offering convenient thickness control due to its structural properties.

**Methods**

**X-ray crystallography.** X-ray diffraction analysis of InGaS$_3$ single crystals was performed on Bruker D8 QUEST diffractometer with Photon III CMOS detector using Mo K*a* radiation (*l* = 0.71073 Å) focused by multilayer Montel mirror. The full dataset was collected at the temperature of 100 K within *ω*-scans indexed with cell-now. It was integrated through SAINT from the SHELXTL PLUS package[42]. Absorption correction was completed by a multiscan approach implemented from SADABS[43]. The crystal structure was solved by direct methods and refined anisotropically with full-matrix $F^2$ least-squares technique using SHELXTL PLUS package. The structure was refined in a chiral P6$_5$ space group with the resulting Flack parameter of 0.06 (2) being close to zero[44]. The second virtual inversion twin-component was added to refinement by TWIN/BASF instruction to account for the absolute structural parameter. This led to a slight reduction of *R*-value from 2.50 % to 2.49 %. CSD reference number 2145523 contains supplementary crystallographic data for this manuscript. It can be obtained free of charge from Cambridge Crystallographic Data Centre *via* www.ccdc.cam.ac.uk/data_request/cif.

**Density functional theory.** Calculations of electronic bandstructure and optical constants were performed using the Vienna ab-initio simulation package VASP[45,46] within generalized gradient approximation (GGA)[47]. The Heyd-Scuseria-Ernzerhof (HSE) hybrid functional[48] in combination with Perdew-Burke-Ernzerhof (PBE)[49] potentials was applied for accurate description of the electronic bandstructure. Electron-ion interactions were described by projector-augmented wave (PAW) method[49,50]. The cutoff energy for the plane-wave basis set was set to 400 eV. The first Brillouin zone of the supercells was sampled with a 6×6×3 Monkhorst-Pack mesh[51] of *k*-points for bulk supercell and 6×6×1 for monolayer case. The maximal force convergence tolerance settings for geometry optimization was set to 0.01 eVÅ$^{-1}$.

**Lattice dynamics calculations.** First principle lattice dynamics calculations were carried out with the CASTEP code package[52]. The crystal structure was fully optimized using LD approximation with CA-PZ exchange-correlation functional based on Ceperley and Alder numerical representation[53] parameterized by Perdew and Zunger[54]. Maximal force and stress tensor convergence tolerance settings for geometry optimization were set to 0.01 eVÅ$^{-1}$ and 0.02 GPa, respectively. $5s_2\ 5p_1$, $3d_{10}\ 4s_2\ 4p_1$, and $3s_2\ 3p_4$ orbitals were considered occupied with valence electrons for Indium, Gallium and Sulphur, respectively. Plane-wave cutoff energy was set to 880 eV for 4×4×2 sampling of the first Brillouin zone by the Monkhorst-Pack method.

**Atomic Force Microscopy.** The thickness and surface morphology of InGaS$_3$ sheets were accurately characterized by an atomic force microscope (NT-MDT Ntegra II) operated in a HybriD mode at ambient conditions. AFM images were acquired using silicon tips from TipsNano (GOLDEN, NSG 10) with spring constant of 11.8 N/m, head curvature radius < 10 nm, and resonant frequency of 240 kHz. The measurements were performed at a scan rate of 0.3 Hz and 512 pixel resolution. The obtained datasets were analyzed by the Gwyddion software.

**Imaging spectroscopic ellipsometry.** To analyze anisotropic optical response of InGaS$_3$, we used commercial imaging spectroscopic ellipsometer Accurion nanofilm_ep4 in the nulling mode. Ellipsometry spectra were recorded for four sheets with various thicknesses (*h* = 3.6 nm, 68.5 nm, 103.0 nm, and 277.4 nm) in the spectral range from ultraviolet (360 nm) to near-infrared (1700 nm). For ellipsometry analysis, we followed the algorithm described in Supplementary Note 2 of our recent work[24].

**Supplementary Information**

Supplementary material contains extra characterization results of non-van der Waals InGaS$_3$ sheets by a variety of techniques: energy-dispersive X-ray analysis, Raman spectroscopy, transmission electron diffraction patterns, AFM scans of inhomogeneous ultrathin sheets. It incorporates the summary data on X-ray diffraction patterns, and includes first-principle evaluation of optical constants along with CASTER calculations of Raman-active modes for recently reported[20] orthorhombic phase of InGaS$_3$.

**Acknowledgements**

This work was supported by Russian Science Foundation (No. 21-79-00218).